
\magnification=1200
\noindent
\nopagenumbers
\vskip5truecm
\centerline{HIGHER-ORDER LAGRANGIAN THEORIES}
\centerline{AND NOETHERIAN SYMMETRIES}
\vskip5truecm
\centerline{D. R. Grigore}
\centerline{ Department of Theoretical Physics}
\centerline{Bucharest-Magurele, Romania}
\vskip5truecm
\centerline{ABSTRACT}

A geometric global formulation of the higher-order
Lagrangian formalism for systems with
finite number of degrees of freedom is provided. The formalism
is applied to the study of systems with groups of Noetherian
symmetries.
\vfill \eject

\pageno=1
\footline{\hss\tenrm\folio\hss}
{\bf 1. Introduction}

This paper is a continuation of [1] in which a generalization of the
usual Lagrangian formalism for first-order Lagrangian systems was
provided. The ideas of [1] were suggested by the reformulation of the
Lagrangian formalism for systems with finite number of degrees
of freedom due to
Poincar\'e, Cartan and Lichnerowicz (see for instance [2])
and Souriau [3]. The main idea is to use for the kinematical
description the first-order jet construction and for the
dynamical description a 2-form $\sigma$ having as characteristic
system exactly the Euler-Lagrange equations.
This formalism is particularly suited for the description of
Lagrangian systems with groups of Noetherian symmetries.

In this paper we will provide a generalization of these ideas to
higher-order Lagrangian systems. As in [1] we will consider the most
general situation in which the basic kinematic manifold $S$ does
not have a canonical fibered structure over the time axis:
$S \rightarrow T$. The case when this fibration exists has been
extensively studied in the literature. We distinguish two
approaches which are however closely related. One line of
argument tries to generalize the Poincar\'e-Cartan 1-form
[4],[5] and another uses a 2-form which has as associated system
the Euler-Lagrange equations [6]-[8]. Our approach is very
similar with the second point of view. Some formal differences with
respect to this approach should be noted. The first one has been
alluded above, namely we do not need the fiber bundle structure
$S \rightarrow T$.
Another difference consists in the choice of
the kinematics. In [6]-[8] one starts form the Euler-Lagrange equations;
if the order of these equations is $s$ then one works on the
$s$-order jet bundle extention of $S$: $J^{s}(S)$. On the other
hand the notion of kinematics is intimatedly connected with the
number of ``initial conditions'' which uniquely determine the
evolution of the system. We will prove that if the ``initial
conditions'' refer to the velocities of order up to $r$, then
one can work on the $r$-order jet bundle extention of $S$: $J^{r}(S)$.
In this way all reference to dynamics dissapears from the
construction of kinematics, which is rather natural.
Moreover, in this way we agree with the usual approach in the
physics literature. For instance, for the case when the
``initial conditions'' refer to the initial position and
velocity ($r = 1$) one usually works on the first-order jet
bundle extension of $S$ (see e.g. [1]). Let us note that in
general $s = r + 1$ so our approach reduces the order of the jet
bundle extension by 1. This simplification can be of some
importance in practical computations.
We are able to prove that the 2-form in our paper and the 2-form
in [6]-[8] are very closely related.

In Section 2 we present the general formalism. We will stress on
the possibility of defining globally the 2-form $\sigma$ in the
same spirit as in [1] and will we give a natural definition
for the notion of regularity closely related with the one in [6]-[8].
We will analyse afterwards Lagrangian
systems with groups of Noetherian symmetries. In the end we will
comment on the relation of our approach with the one in [6]-[8].

In Section 3 we illustrate the usefulness of the framework for
the cases $r = 2$ and $r = 3$ with Galilei and Poincar\'e
invariance as in [1].

In Section 4 we analyse again the case of Poincar\'e invariance
but we use a homogeneous formalism in the same spirit as in [9],
[10].

\vskip 2truecm
{\bf 2. General Formalism for Higher-Order Lagrangian Systems}

2.1 We emphasise again that we will consider only the case of
finite number of degrees of freedom.
Suppose
$S$
is a
$(N+1)$-dimensional
manifold interpreted as the ''space-time'' manifold of the
system. $S$ will be called the {\it kinematical manifold} of
the system. We remind briefly how one can constuct the
$s$-order jet bundle extension of $S$.
One starts from the projective tangent bundle:
$$
J^{1}_{1}(S) \equiv PT(S)
$$
with the canonical projection
$\pi_{0}: J^{1}_{1}(S) \rightarrow S$.
For
$k > 0$,
$J^{k + 1}_{1}(S)$
will be a bundle over
$J^{k}_{1}(S)$: $\pi_{k}: J^{k + 1}_{1}(S) \rightarrow
J^{k}_{1}(S)$.
If
$x_{k} \in J^{k}_{1}(S)$
then the fiber over
$x_{k}$
consists of the straight lines of
$T_{x_{k}}(J^{k}_{1}(S))$
which projects into elements of
$J^{k}_{1}(S)$ i.e.
$$
J^{k + 1}_{1}(S)_{x_{k}} = \{ R\cdot v \vert
v \in T_{x_{k}}(J^{k}_{1}(S)), (\pi_{k})_{*}v \in
J^{k}_{1}(S)_{\pi_{k - 1}(x_{k})} \}.\eqno(2.1)
$$

We need a convenient system of local charts on
$J^{r}_{1}(S)$.
Suppose
$(t,q^{A}_{0}), A =1,...,N$
is a local system of coordinates on the open set
$U_{0} \subset S$.
Then one proves by recurrence that a system of charts on
$J^{k}_{1}(S), k = 1,...,r$
exists and it is determned as follows:

1) one has a system of open sets
$U_{k} \subset J^{k}_{1}(S), k = 1,...,r$
such that
$$
U_{k+1} \subseteq (\pi_{k})^{-1}(U_{k}),~~~ k = 0,...,r-1;
$$

2) on
$U_{k}$
one has local coordinates
$(t,q^{A}_{0},...,q^{A}_{k}), A =1,...,N$
such that if the coordinates of
$x_{k} \in J^{k}_{1}(S)$
are
$(t,(q^{A}_{0})_{0},...,(q^{A}_{k})_{0})$,
then the straight line in
$T_{x_{k}}(J^{k}_{1}(S))$
corresponding to
$(t,(q^{A}_{0})_{0},...,(q^{A}_{k+1})_{0})$
is generated by the vector:
$$
\left( { \delta \over \delta t}\right)_{k} \equiv
\left( {\partial \over \partial t} \right)_{x_{k}}+
\sum_{i=0}^{k} (q^{A}_{i+1})_{0}
\left( {\partial \over \partial q^{A}_{i}}\right)_{x_{k}}.\eqno(2.2)
$$
Here
$k= 0,...,r - 1$
and the summation convention over the dummy indices is used.

Indeed, one has the relationship:
$$
(\pi_{k})_{*} \left( {\delta \over \delta t}\right)_{k} =
\left( {\delta \over \delta t}\right)_{k-1},\eqno(2.3)
$$
and the assertion follows from the definition (2.1).

It is clear that
$q^{A}_{1}$
we will interpreted as the coordinates of the
velocity,
$q^{A}_{2}$
as the coordinates of the accelerations (i.e. velocities of
order 2) and so on.

Then, by definition, an {\it evolution space} over
$S$
is any open subbundle
$E$
of the bundle
$J^{r}_{1}(S)$.

2.2 Let
$E \subseteq J^{r}_{1}(S)$ be an evolution space over
$S$.
We define:
$$
\wedge_{LS}(E) \equiv \{ \sigma \in \wedge^{2}(E) \vert
i_{V_{1}}i_{V_{2}} \sigma = 0, \forall V_{i} \in Vect(E)
{}~~s.t.~~ (\pi_{r})_{*}V_{i} =0~~
i = 1,2 \}.\eqno(2.4)
$$

It is clear that in the local coordinates above,
$
\forall \sigma \in \wedge_{LS}(E)
$
has the expression:
$$
\sigma = \sum_{i=0}^{r-1}\sigma^{i}_{AB} dq^{A}_{r}
\wedge \delta q^{B}_{i} +
\sigma_{A} dq^{A}_{r} \wedge dt
+ {1 \over 2} \sum_{i,j=0}^{r-1} \tau_{AB}^{ij}
\delta q^{A}_{i} \wedge \delta q^{B}_{j}
+ \sum_{i=0}^{r-1} \tau_{A}^{i} \delta q^{A}_{i} \wedge dt.\eqno(2.5)
$$
Here we have denoted:
$$
\delta q^{A}_{i} \equiv d q^{A}_{i} - q^{A}_{i+1} dt
{}~~~(i = 0,...,r-1),\eqno(2.6)
$$
and we can suppose that:
$$
\tau_{AB}^{ij} = - \tau_{BA}^{ji}.\eqno(2.7)
$$

Now we have a key result:

{\bf Proposition 1:} In the notations (2.5) the equations:
$$
\sigma_{A} = 0\eqno(2.8)
$$
$$
\sigma^{i}_{AB} = 0~~~(i = 1,...,r -1)\eqno(2.9)
$$
$$
\tau^{i}_{A} = 0~~~(i = 1,...,r - 1)\eqno(2.10)
$$
are globally defined.

{\bf Proof:} Let
$
U_{0}, \bar{U}_{0}\in S
$
with local coordinates
$
(t,q^{A}_{0})
$
and
$
(\bar{t},\bar{q}^{A}_{0})
$
respectively. Let
$
(t,q^{A}_{0},...,q^{A}_{k})
$
and
$
(\bar{t},\bar{q}^{A}_{0},...,\bar{q}^{A}_{k})
$
the local systems of coordinates on
$
J^{r}_{1}(S)
$
constructed as in Subsection 2.1. Let us suppose that
$
U_{0} \cap \bar{U}_{0} \not= \emptyset
$
and that the change of charts has the form:
$$
\bar{t} = f(t,q);~~~\bar{q}^{A}_{0} = g^{A}_{0}(t,q)
.\eqno(2.11)
$$

We must show that:
$$
\sigma_{A} = 0,~~\sigma^{i}_{AB} = 0,~~\tau^{i}_{A} = 0~~~
(i = 1,...,r - 1)
$$
{\it iff}
$$
\bar{\sigma}_{A} = 0,~~\bar{\sigma}^{i}_{AB} = 0,~~
\bar{\tau}^{i}_{A} = 0~~~(i = 1,...,r - 1).
$$
In fact it is sufficient to prove only  one implication. First one needs
the change of charts induced on
$
J^{r}_{1}(S),
$
i.e. the functions
$
\bar{q}^{A}_{i}~~~(i = 1,...,r).
$
One can show by induction that one has:
$$
\bar{q}^{A}_{i} = g^{A}_{i}(t,q_{1},...,q_{i})~~~(i = 1,...,r).\eqno(2.12)
$$

The functions
$
g^{A}_{i}
$
are determined recurrsively from the formulae:
$$
g^{A}_{k+1} = \left( {\delta f \over \delta t}\right)^{-1}
{}~{\delta g^{A}_{k} \over \delta t}~~~(k = 0,..., r -
1),\eqno(2.13)
$$
where:
$$
{\delta \over \delta t} \equiv \left( {\delta \over \delta t}
\right) _{r-1} = {\partial \over \partial t} + \sum_{k=0}^{r-1}
q^{A}_{k+1} {\partial \over \partial q^{A}_{k}}.\eqno(2.14)
$$

{}From (2.13) one determines that
$
\delta\bar{q}^{A}_{i}~~(i = 0,...,r - 1)
$
has the following structure:
$$
\delta\bar{q}^{A}_{i} = \sum_{j=0}^{i} g^{jA}_{iB} \delta q^{B}_{j}
$$
for some functions
$
g^{jA}_{iB}.
$
Now the desired implication follows. Q.E.D.

2.3 By definition, a {\it Lagrange-Souriau form} on the
evolution space
$
E \subseteq J^{r}_{1}(S)
$
is any closed 2-form
$
\sigma \in \wedge_{LS}(E)
$
verifying (2.8) - (2.10). If we take into account these
relations, the expression (2.5) simplifies to:
$$
\sigma = \sigma_{AB} dq^{A}_{r}
\wedge \delta q^{B}_{0} +
{1 \over 2} \sum_{i,j=0}^{r-1} \tau_{AB}^{ij}
\delta q^{A}_{i} \wedge \delta q^{B}_{j}
+ \tau_{A} \delta q^{A}_{0} \wedge dt.\eqno(2.15)
$$

In practical computations we will need the explicit content of closedness
condition:
$$
d\sigma = 0.\eqno(2.16)
$$

The computations are rather tedious but elemenary so we provide only the final
result. We have two distint cases. For
$
r = 1
$
we get, as expected, the results obtained in [1] (namely eqs.
(2.10)-(2.15) from [1]).

The case
$
r > 1
$
in which we are interested in this paper is a litttle more
complicated; we get:
$$
{\partial \sigma_{AB} \over \partial q^{C}_{r}} = 0.\eqno(2.17)
$$
$$
{\delta\sigma_{AB} \over \delta t} - {\partial \tau_{A} \over
\partial q^{B}_{r}} + \tau^{0,r-1}_{AB} = 0.\eqno(2.18)
$$
$$
\tau^{1,r-1}_{AB} = \sigma_{BA}.\eqno(2.19)
$$
$$
\tau^{r-1,i}_{AB} = 0~~~(i = 2,...,r - 1).\eqno(2.20)
$$
$$
{\partial \sigma_{AB} \over \partial q^{C}_{0}} -
{\partial \sigma_{AC} \over \partial q^{B}_{0}} +
{\partial \tau^{00}_{BC} \over \partial q^{A}_{r}} = 0.\eqno(2.21)
$$
$$
{\partial \sigma_{AB} \over \partial q^{C}_{i}} +
{\partial \tau^{0i}_{BC} \over \partial q^{A}_{r}} = 0~~~(i =
1,...,r - 1).\eqno(2.22)
$$
$$
{\partial \tau^{ij}_{AB} \over \partial q^{C}_{r}} = 0~~~
(i,j = 1,...,r - 1).\eqno(2.23)
$$
$$
{\delta \tau^{00}_{AB} \over \delta t} +
{\partial \tau_{B} \over \partial q^{A}_{0}} -
{\partial \tau_{A} \over \partial q^{B}_{0}} = 0.\eqno(2.24)
$$
$$
{\delta \tau^{0i}_{AB} \over \delta t} +
\tau^{0,i-1}_{AB} - {\partial \tau_{A} \over \partial q^{B}_{i}}
= 0~~~(i = 1,...,r - 1).\eqno(2.25)
$$
$$
{\delta \tau^{ij}_{AB} \over \delta t} +
\tau^{i-1,j}_{AB} - \tau^{j-1,i}_{BA} = 0~~~(i,j = 1,...,r - 1).\eqno(2.26)
$$
$$
{\partial \tau^{ij}_{AB} \over \partial q^{C}_{k}} +
{\partial \tau^{jk}_{BC} \over \partial q^{A}_{i}} +
{\partial \tau^{ki}_{CA} \over \partial q^{B}_{j}} = 0~~~(i,j,k = 0,...,r
-1).\eqno(2.27)
$$
with the convention that equation (2.20) disapears for
$r =2$.

We will call (2.17)-(2.27) the {\it structure equations}.

2.4 A {\it Lagrangian system} on
$S$
is, by definition, any couple
$
(E,\sigma)
$
where
$
E \subseteq J^{r}_{1}(S)
$
is an evolution space over
$S$
and
$\sigma$
is a Lagrange-Souriau form on
$E$.

We will give now the main concepts involved in the study of
Lagrangian systems. There is a close resemblance with [1].

Two Lagrangian systems
$
(E_{1}, \sigma_{1})
$
and
$
(E_{2}, \sigma_{2})
$
on
$S$
are called {\it equivalent} if there exists a diffeormorphism
$
\phi \in Diff(S)
$
such that
$
\dot\phi(E_{1}) = E_{2}
$
and:
$$
(\dot\phi)^{*} \sigma_{2} = \sigma_{1}.\eqno(2.28)
$$

Here
$
\dot\phi \in Diff(J^{r}_{1}(S))
$
is the natural lift of
$\phi$.

To introduce the notion of non-degeneracy we need:

{\bf Proposition 2:} Let
$
\sigma
$
be a Lagrange-Souriau form written locally as in (2.15). Then
the condition:
$$
det(\sigma_{AB}) \not= 0,\eqno(2.29)
$$
is globally defined.

{\bf Proof:} Using the expressions of the change of charts
$
(t,q^{A}_{0},...,q^{A}_{r}) \rightarrow (f,g^{A}_{0},...,g^{A}_{r})
$
form Proposition 1, it is rather easy to get the transformation
law for the functions
$
\sigma_{AB}
$:
$$
\sigma_{AB} = \bar{\sigma}_{CD} {J^{C}}_{A} {J^{D}}_{B} \left(
{\delta f\over \delta t}\right)^{-r},
$$
where
$${J^{A}}_{B} = {\partial g^{A}_{0} \over \partial q^{B}_{0}} -
g^{A}_{1} {\partial f \over \partial q^{B}_{0}}
$$

So, it remains to show that the inversability of the change of
charts implies:
$$
det(J) \not= 0
$$
which is not very complicated. Q.E.D.

We say that the Lagrangian system $(E,\sigma)$ is {\it non-degenerated}
if the condition (2.29) is true.

An {\it evolution} is any immersion
$
\gamma: T \rightarrow S
$,
where
$T$
is some one-dimensional manifold (the ``time'' manifold).
Usually
$T = R$.
If
$\gamma$
is an evolution and
$
\dot\gamma:T \rightarrow J^{r}_{1}(S)
$
is the natural lift of
$\gamma$,
we say that
$\gamma$
verifies the {\it Euler-Lagrange equations of motion} if:
$$
(\dot\gamma)^{*} i_{Z} \sigma = 0.\eqno(2.30)
$$
for any vector field $Z$ on $E$.

It can be proven that if
$\sigma$
is exihibited in the form (2.15) then the Euler-Lagrange
equations are:
$$
\sigma_{BA}\circ\dot\gamma~{d^{r+1} x^{B} \over d
t^{r+1}} - \tau_{A}\circ\dot\gamma = 0.\eqno(2.31)
$$
where
$\gamma$
is chosen of the form
$
\gamma: t \mapsto (t,x^{A}(t)).
$
So, it is clear that these equations are, in general, of order
$s = r + 1$.
Moreover if
$
(E,\sigma)
$
is a non-degenerated Lagrangian system, then it follows that
$\gamma$
is determined by the ``initial condition''
$
x^{A}_{0}(t_{0}),...,x^{A}_{r}(t_{0})
$
(at least locally). So, the non-degeneracy condition expresses
in an abstract form the notion of (Newtonian) determinism.

Usually, one has a {\it causality relationship} on
$S$
(i.e. an order relation on
$S$)
and then it is natural to require that the points in
$
Im(\gamma)
$
are in a causality relationship.

We also note that the phase space can be constructed following
Souriau idea [3]. Namely if
$
dim(Ker(\sigma))
$
is constant on
$E$,
the {\it phase space} is the characteristic foliation of
$
(E,\sigma)
$.

We close this Subsection with the formulation of the notion of
(Noetherian) symmetriy.
Let
$
(E,\sigma)
$
be a Lagrangian system on
$S$.
A {\it symmetry} of
$
(E,\sigma)
$
is any diffeomorphism
$
\phi \in Diff(S)
$
such that: (a)
$
\dot\phi(E) = E
$;
(b) if
$\gamma$
verifies the Euler-Lagrange equations, then
$
\phi \circ \gamma
$
is also a solution of these equations.

If
$
\phi \in Diff(S)
$
verifies:
$$
(\dot\phi)^{*} \sigma = \sigma \eqno(2.32)
$$
then, it is easy to prove that
$\phi$
is a symmetry of the Lagrangian system. We call such type of symmetries,
{\it Noetherian symmetries}. A {\it Noetherian group of symmetries}
is an action of a group on
$S$:
$
G \ni g \rightarrow \phi_{g}\in Diff(S)
$
such that any
$
\phi_{g}
$
is a Noetherian symmetry, i.e.:
$$
(\dot\phi_{g})^{*} \sigma = \sigma~~~(\forall g \in G).\eqno(2.33)
$$

Let us remark that in this case one must apropriately modify the
notion of equivalence for Lagrangian systems (requiring that
$\phi$
in (2.28) is a
$G$-morphism)
and the notion of causality (requiring that the order relation on
$S$
is
$G$-invariant).

As illustrated in [1], this formulation of the Lagrangian
formalism allows an explicit  classification of all Lagrangian
systems
$
(E,\sigma)
$
on
$S$
having a certain group
$G$
of Noetherian symmetries, for many groups
$G$
of physical interest.

It is noteworthy that the entire structure
induces the well-known symplectic formalism on the
associated phase space [3].

2.5 Now we connect the abstract scheme developped above with the usual
Lagrangian formalism. The way to do this is to first  prove

{\bf Proposition 3:} Let
$(E,\sigma)$
be a Lagrangian system over
$S$.
Then the 2-form
$\sigma$
can be written locally as follows:
$$
\sigma = d\theta
\eqno(2.34)
$$
where
$\theta$
has the following expression:
$$
\theta = L dt + \sum_{i=0}^{r-1} f^{i+1}_{A} \delta q^{A}_{i}.\eqno(2.35)
$$
Here
$L$
is a local function of
$t$
and
$q^{A}_{i} (i = 0,...,r)$. We have denoted
$$
f^{i+1}_{A} \equiv \sum_{k=0}^{r-1-i} (-1)^{k} \left( {\delta \over
\delta t}\right)^{k} {\partial L \over \partial q^{A}_{i+1+k}}\eqno(2.36)
$$
for
$i = 0,...,r -1$. Moreover, the following identities must be obeyed:
$$
{\partial f^{i+1}_{A} \over \partial q^{B}_{r}} = 0
{}~~~(i = 1,...,r -1) \eqno(2.37)
$$
in the case
$r \geq 2$.

{\bf Proof:} From the closedness condition (2.16) we have
locally (2.34) where
$\theta$
is generically of the form:
$$
\theta = L dt + \sum_{i=0}^{r-1} f^{i+1}_{A} \delta q^{A}_{i}
+ \theta_{A} d q^{A}_{r}.
$$

So, we have locally:
$$
\sigma = d\theta = {\partial \theta_{A} \over \partial q^{B}_{r}}
d q^{B}_{r} \wedge d q^{A}_{r} + \cdots
$$
where by
$\cdots$
we understand contributions from
$\Lambda_{LS}(E)$.
Because
$\sigma \in \Lambda_{LS}(E)$
by definition, we must have:
$
{\partial \theta_{A} \over \partial q^{B}_{r}} =
{\partial \theta_{B} \over \partial q^{A}_{r}}
$
so,
$\theta_{A}$
is of the form:
$
\theta_{A} = {\partial f \over \partial q^{A}_{r}}.
$

If one redefines
$
\theta \rightarrow \theta - d f
$,
then (2.34) stays true, but the last contribution in the
expression above of
$\theta$
dissapears. It means that
$\theta$
aquires the expression (2.35).

Next, one imposes the conditions (2.8)-(2.10) from the statement
of Proposition 1 and gets besides (2.37) the following
relations:
$$
{\partial L \over \partial q^{A}_{r}} - f^{r}_{A} = 0\eqno(2.38)
$$
and
$$
{\partial L \over \partial q^{A}_{i}}
- {\delta f^{i+1}_{A} \over \delta t}
- f^{i}_{A} = 0~~~(i = 1,...,r -1)\eqno(2.39)
$$
which can be used to prove recurrsively (2.36). Q.E.D.

We will call
$L$
a (local) {\it Lagrangian}. If
$\theta$
is of the form (2.35) we denote it by
$\theta_{L}$;
we also denote
$
\sigma_{L} \equiv d\theta_{L}.
$

Now one recovers the usual Euler-Lagrange equations. Indeed, if
$\sigma = \sigma_{L}$
one easily shows that (2.31) becomes:
$$
\sum_{k=0}^{r} (-1)^{k} {d^{k} \over d t^{k}}\left( {\partial L \over
\partial q^{A}_{k}} \circ\dot\gamma \right)= 0.\eqno(2.40)
$$

The identities (2.37) restrict the form of the Lagrangian
$L$
and can be used to ``lower'' the order of this function, i.e. to
choose it independent of
$q^{A}_{i}$
for some
$i \geq c +1$
where
$c \leq  r$.
(This makes sense for
$r \geq 2$).
Indeed, we have:

{\bf Proposition 4:} Let
$
(E,\sigma)
$
be a Lagrangian system over
$S$,
$
E \subseteq J^{r}_{1}(S)
$
$(r \geq 2)$.
Define:
$$
c \equiv \left[ {r \over 2} \right] + 1.\eqno(2.41)
$$

Then there exists a local Lagrangian
$L_{min}$
such that:
$$
{\partial L_{min} \over \partial q^{A}_{i}} = 0
{}~~~(i = c + 1,...,r)\eqno(2.42)
$$
and instead of (2.35) and (2.36) we have:
$$
\theta = L_{min} dt + \sum_{i=0}^{c-1} f^{i+1}_{A} \delta q^{A}_{i}
\eqno(2.43)
$$
and
$$
f^{i+1}_{A} \equiv \sum_{k=0}^{c-1-i} (-1)^{k} \left( {\delta \over
\delta t}\right)^{k} {\partial L \over \partial q^{A}_{i+1+k}}.\eqno(2.44)
$$

Moreover in the case
$r = 2c$,
$L_{min}$
is linear in
$q_{c}$:
$$
{\partial^{2} L_{min} \over \partial q^{A}_{c} \partial q^{B}_{c}}
= 0 .\eqno(2.45)
$$

{\bf Proof:} For
$r = 2$
we have
$c = 2 = r - 1$
so (2.35) and (2.36) coincide with (2.43) and (2.44)
respectively. Moreover (2.37) and (2.38) give (2.45).

Suppose now that
$r \geq 2$.
{}From (2.38) it follows that
$L$
has the form:
$$
L = q^{A}_{r} f^{r}_{A} + L'
$$
where
$L'$
does not depend on
$q_{r}$.
We insert this expression into (2.39) for
$i = r - 1$
and differentiate with respect to
$q^{B}_{r}$;
taking into account that
$r \geq 2$
we have (2.37) and we get:
$
{\partial f^{r}_{B} \over \partial q^{A}_{r-1}} =
{\partial f^{r}_{A} \over \partial q^{B}_{r-1}}
$
so
$f^{r}_{A}$
is of the following form:
$
f^{r}_{A} = {\partial f^{r} \over \partial q^{A}_{r-1}}
$
for some function
$f^{r}$
which does not depend on
$q_{r}$.
If we redefine
$
\theta \rightarrow \theta -d f^{r}
$,
then (2.34) stays true, but instead of  (2.35) we have,
up to some redefinitions:
$$
\theta = L dt + \sum_{i=0}^{r-2} f^{i+1}_{A} \delta q^{A}_{i}
$$

Now one iterates the procedure; the recurrence stops when
$\theta$
aquires the expression (2.43) with
$L_{min}$
verifying (2.42). Imposing again the conditions (2.8)-(2.10) we
get (2.37) and (2.44). Now it is rather easy to show that if
$r$
is odd then (2.37) are identically verified. If
$r$
is even then (2.37) are identically verified for
$i \geq 1$
and for
$i = 1$
we obtain exactly (2.45). Q.E.D.

We will call
$L_{min}$
a {\it minimal Lagrangian} and
$c$
the {\it minimal order}.

One also has the following consequence of the result above:

{\bf Proposition 5:} The non-degeneracy condition has the
following local formulation:

- for
$r$
odd:
$$det\left( {\partial^{2} L_{min} \over \partial q^{A}_{c}
\partial q^{B}_{c}}\right) \not= 0\eqno(2.46)
$$

- for
$r$
even:
$$
det \left( {\partial^{2} L_{min} \over \partial q^{A}_{c}
\partial q^{B-1}_{c-1}} - A \leftrightarrow B \right) \not= 0.\eqno(2.47)
$$

{\bf Proof:} Consists in the computation of the functions
$
\sigma_{AB}
$ in terms of
$L_{min}$.
Q.E.D.

One should compare Propositions 3-5 above with the similar
results obtained in [7].

2.6 Next, one establishes that working with
$\sigma$
instead of
$L$
eliminates from the game the so-called total derivative
Lagrangians. Indeed we have:

{\bf Proposition 6:} The following three assertions are equivalent:

(a)
$\sigma_{L} = 0;$

(b) the Euler-Lagrange equations for
$
(E,\sigma_{L})
$
are trivial;

(c)
$L$
is of the form
$$
L = {\delta \Lambda \over \delta t}.\eqno(2.48)
$$
with
$\Lambda$
a function of
$t$
and
$
q^{A}_{i}~~~(i = 0,...,r - 1).
$

{\bf Proof:} The implication
$(a) \Longleftrightarrow (c)$
follows easily from the closedness condition (2.16). The
implication
$(b) \Longleftrightarrow (c)$
follows from (2.39) in a standard way by induction. Q.E.D.

Lagrangians of the type (2.48) are called {\it total derivative}
Lagrangians.

Finally, we connect with the usual definition of the Noetherian
symmetries. Indeed, suppose that the transformation
$\phi$
in (2.32) has the following form in local coordinates:
$
\phi(t,q_{0}) = (f(t,q_{0}),g_{0}(t,q_{0}))
$
and
$\sigma = \sigma_{L}$.
Then one easily establish using Propositions 1 and 5 that
$L$
verifies the following identity:
$$
{\delta f\over \delta t} L\circ\dot\phi - L = {\delta \Lambda
\over \delta t}.\eqno(2.49)
$$

Let us suppose that
$\theta$
is globally defined on
$E$.
Then one can introduce the {\it action functional}:
$$
A(\gamma) \equiv \int (\dot\gamma)^{*} \theta.\eqno(2.50)
$$

Then the solutions of the Euler-Lagrange equations are extremals
of this functional and moreover (2.49)
is equivalent to the usual definition
of Noetherian symmetries:
$$
A(\phi \circ \gamma) = A(\gamma) + a~~trivial~~action.\eqno(2.51)
$$
where by a {\it trivial action} we mean an action giving trivial
Euler-Lagrange equations of motions. (Such actions are also called
boundary terms).

We have recovered the whole structure of the Lagrangian formalism.
However, we did not need a configuration space like in the usual
formulations.

Let us stress once again that the concept of  Lagrangian is
only a local one. Even if we can take
$U = E$,
$L$
remains a chart dependent object and cannot be considered, in general,
as a function on
$E$.

2.7 We close this Section with some comments regarding the
relation of our approach with the one in [6]-[8]. As stated in
the Introduction, in this approach the kinematics takes place in
$
J^{s}_{1}(S) = J^{r+1}_{1}(S)
$.

Instead of (2.15) one takes
$\sigma$
of the form:
$$
\sigma = E_{A} \delta q^{A}_{0} \wedge d t + {1\over 2}
\sum_{i,j=0}^{r} F^{ij}_{AB} \delta q^{A}_{i} \wedge
\delta q^{A}_{j}\eqno(2.52)
$$
where we have implicitely assumed that (2.6) has been extended
for
$i = r$
also (which is possible in
$
J^{r+1}_{1}(S)
$).
Inspecting all the results of Subsections 2.2-2.6 we conclude
that they stay true in this case also (with some apropriate
modifications). For instance the global definition of
$\sigma$
involves the following conditions:

(a)
$i_{V} \sigma = 0
$
for any
$
V \in Vect(J^{r+1}_{1}(S))
$
such that
$
(\pi_{r+1})_{*} V = 0
$.

(b)
$\sigma$
does not contain the differential forms
$
\delta q^{A}_{i} \wedge d t~~~(i = 1,...,r).
$

In rest, one must take care to replace everywhere
$
{\delta \over \delta t} \equiv \left( {\delta \over \delta t}
\right)_{r-1}
$
with
$
{\delta \over \delta t} \equiv \left( {\delta \over \delta t}
\right)_{r}.
$

A final observation is the following one. By some combinatorics
one can eliminate completely the variables
$q_{r+1}$
from the expression (2.52) and one obtains exactly the
expression (2.15). So, (2.52) is the pull-back of (2.15).
\vskip 1truecm
{\bf 3. Lagrangian Systems with Groups of Noetherian Symmetries}

A. We consider first the case
$r = 2$.
The kinematic manifold will be taken as
$
S = R \times R^{n-1}
$
with global coordinates
$
(t,q^{A}_{0})~~~(A = 1,...,n - 1)
$.
Then
$
J^{2}_{1} \cong R \times R^{n-1} \times R^{n-1}
$
with global coordinates
$
(t,q^{A}_{0},q^{A}_{1},q^{A}_{2})~~~(A = 1,...,n - 1)
$.

3A.1 In this case the closedness conditions (2.17)-2.27) reduces
to:
$$
{\partial \sigma_{AB} \over \partial q^{C}_{2}} = 0.\eqno(3.1)
$$
$$
{\delta\sigma_{AB} \over \delta t} - {\partial \tau_{A} \over
\partial q^{B}_{2}} + \tau^{0,1}_{AB} = 0.\eqno(3.2)
$$
$$
\tau^{1,1}_{AB} = \sigma_{BA}.\eqno(3.3)
$$
$$
{\partial \sigma_{AB} \over \partial q^{C}_{0}} -
{\partial \sigma_{AC} \over \partial q^{B}_{0}} +
{\partial \tau^{00}_{BC} \over \partial q^{A}_{2}} = 0.\eqno(3.4)
$$
$$
{\partial \sigma_{AB} \over \partial q^{C}_{1}} +
{\partial \tau^{01}_{BC} \over \partial q^{A}_{2}} = 0 .\eqno(3.5)
$$
$$
{\partial \tau^{11}_{AB} \over \partial q^{C}_{2}} = 0 .\eqno(3.6)
$$
$$
{\delta \tau^{00}_{AB} \over \delta t} +
{\partial \tau_{B} \over \partial q^{A}_{0}} -
{\partial \tau_{A} \over \partial q^{B}_{0}} = 0.\eqno(3.7)
$$
$$
{\delta \tau^{01}_{AB} \over \delta t} +
\tau^{00}_{AB} - {\partial \tau_{A} \over \partial q^{B}_{1}} = 0.\eqno(3.8)
$$
$$
{\delta \tau^{11}_{AB} \over \delta t} +
\tau^{01}_{AB} - \tau^{01}_{BA} = 0.\eqno(3.9)
$$
$$
{\partial \tau^{ij}_{AB} \over \partial q^{A}_{k}} +
{\partial \tau^{jk}_{BC} \over \partial q^{A}_{i}} +
{\partial \tau^{ki}_{CA} \over \partial q^{B}_{j}} = 0
{}~~~(i,j,k = 0,1) .\eqno(3.10)
$$

3A.2 We first study the functions
$\sigma_{AB}$.
It is not hard to prove from the relations above that one has
the antisymmetry property:
$$
\sigma_{AB} = - \sigma_{BA}\eqno(3.11)
$$
and also:
$$
{\partial^{2} \sigma_{AB} \over \partial q^{C}_{1} \partial
q^{D}_{1}} = 0.\eqno(3.12)
$$

3A.3 We postulate now invariance  with respect to
spatio-temporal translations and rotations; the action of these
transformations on
$S$
is:
$$
\phi_{R,\eta,a} (t,q_{0}) = (t, + \eta, R q_{0} + a).\eqno(3.13)
$$
Here
$
\eta \in R
$,
$
a \in R^{n-1}
$ and
$
R \in SO(n - 1).
$
The lift of
$
\phi_{R,\eta,a}
$
to
$J^{2}_{1}(S)$
is:
$$
\dot\phi_{R,\eta,a}(t,q_{0},q_{1},q_{2}) = (t+\eta,Rq_{0}+a,Rq_{1},Rq_{2})
.\eqno(3.14)
$$

We require that:
$$
(\dot\phi_{R,\eta,a})^{*} \sigma = \sigma.\eqno(3.15)
$$

It is easy to see that (3.15) is equivalent to the independence
of the functions
$\sigma_{AB}$,
$\tau^{ij}_{AB}$
and
$\tau_{A}$
of the variables
$t$
and
$q_{0}$
and with their rotation covariance.

In particular, if we take into account the relation (3.1) it
follows that
$\sigma_{AB}$
depends only of the variable
$q_{1}$.
Then (3.12) shows that
$\sigma_{AB}$
is a polynomial of the form:
$$
\sigma_{AB} = \sigma_{ABC} q^{C}_{1} + \sigma_{AB}'
$$
with
$\sigma_{ABC}$
and
$\sigma_{AB}'$
some constants.

Because of the rotation covarianvce, the antisymmetry
property (3.11) and (3.10) (for
$i = j = k = 1$),
$\sigma_{AB}$
can be non-zero {\it iff}
$n = 3$;
in this case we have:
$$
\sigma_{AB} = \kappa \varepsilon_{AB}.\eqno(3.16)
$$

{}From (3.17) it follows that:
$$\tau_{AB} = - \kappa \varepsilon_{AB}.\eqno(3.17)
$$

We still have to determine the functions
$\tau^{00}_{AB}$,
$\tau^{01}_{AB}$
and
$\tau_{A}$.
This is not very difficult.One obtains from the closedness
condition that
$\tau^{00}_{AB}$
is constant,
$\tau^{01}_{AB}$
depends only of
$q_{1}$
and verifies
$
{\partial \tau^{01}_{AB} \over \partial q^{C}_{1}} = B
\leftrightarrow C
$,
and
$\tau_{A}$
is given by:
$$
\tau_{A} = \tau^{01}_{AB} q^{B}_{2} + \tau^{00}_{AB} q^{B}_{1} +
t_{A}
$$
with
$t_{A}$
constants. Then Lorentz covariance fixes:
$$
\tau^{00}_{AB} = \lambda \varepsilon_{AB}~~~(\lambda \in R).\eqno(3.18)
$$
$$
\tau^{01}_{AB} = \delta_{AB} f(q_{1}^{2}) + 2 q_{1A} q_{1B}
f'(q_{1}^{2}) .\eqno(3.19)
$$
and
$$
\tau_{A} = \tau^{01}_{AB} q^{B}_{2} + \tau^{00}_{AB} q^{B}_{1}.\eqno(3.20)
$$

Here
$f$
is an arbitrary smooth function.

The solution to our problem is given by (3.16)-(3.20).

It is easy to prove that
$
\sigma = \sigma_{L}
$
with
$L$
of the form
$$
L(q_{1},q_{2}) = - {1 \over 2} \kappa \varepsilon_{AB} q_{1}^{A}
q^{B}_{2} + l(q_{1}^{2}) + {1 \over 2} \lambda \varepsilon_{AB}
q^{A}_{0} q_{1}^{B}\eqno(3.21)
$$
with
$l$
such that
$
f = -2 l'
$.

We stress once again that the only physically intersting case in
the non-degenerared one which occurs only for
$\kappa \not= 0$.

{\bf Remark} If we impose in addition Galilei invariance it can
be proved as in [1] that
$l$
can be chosen of the form
$
l(q_{1}^{2}) = {1 \over 2} m q^{2}_{1}
$
with
$m \in R$
and
$\lambda = 0$.

If we impose in addition the invariance with respect to pure
Lorentz transformations we must have
$\kappa = 0$
which contradicts the non-degeneracy.

B. Next we consider the case
$r = 3$.
We take the same kinematic manifold
$S$
as before
and we identify
$
J^{3}_{1}(S) \cong R \times R^{n-1} \times R^{n-1} \times R^{n-1}
$
with coordinates
$
(t,q_{0},q_{1},q_{2},q_{3})
$.
We will consider as before Galilei and Poincar\'e invariance.

3B.1 In this case the closedness conditions (2.17)-(2.27) become:
$$
{\partial \sigma_{AB} \over \partial q^{C}_{3}} = 0.\eqno(3.22)
$$
$$
{\delta\sigma_{AB} \over \delta t} - {\partial \tau_{A} \over
\partial q^{B}_{3}} + \tau^{02}_{AB} = 0.\eqno(3.23)
$$
$$
\tau^{12}_{AB} = \sigma_{BA}.\eqno(3.24)
$$
$$
\tau^{22}_{AB} = 0.\eqno(3.25)
$$
$$
{\partial \sigma_{AB} \over \partial q^{C}_{0}} -
{\partial \sigma_{AC} \over \partial q^{B}_{0}} +
{\partial \tau^{00}_{BC} \over \partial q^{A}_{3}} = 0.\eqno(3.26)
$$
$$
{\partial \sigma_{AB} \over \partial q^{C}_{1}} +
{\partial \tau^{01}_{BC} \over \partial q^{A}_{3}} = 0 .\eqno(3.27)
$$
$$
{\partial \sigma_{AB} \over \partial q^{C}_{2}} +
{\partial \tau^{02}_{BC} \over \partial q^{A}_{3}} = 0 .\eqno(3.28)
$$
$$
{\partial \tau^{ij}_{AB} \over \partial q^{C}_{r}} = 0~~~
(i,j = 1,2).\eqno(3.29)
$$
$$
{\delta \tau^{00}_{AB} \over \delta t} +
{\partial \tau_{B} \over \partial q^{A}_{0}} -
{\partial \tau_{A} \over \partial q^{B}_{0}} = 0.\eqno(3.30)
$$
$$
{\delta \tau^{01}_{AB} \over \delta t} +
\tau^{00}_{AB} - {\partial \tau_{A} \over \partial q^{B}_{1}}= 0.\eqno(3.31)
$$
$$
{\delta \tau^{02}_{AB} \over \delta t} +
\tau^{01}_{AB} - {\partial \tau_{A} \over \partial q^{B}_{2}}= 0.\eqno(3.32)
$$
$$
{\delta \tau^{ij}_{AB} \over \delta t} +
\tau^{i-1,j}_{AB} - \tau^{j-1,i}_{BA} = 0~~~(i,j = 1,2).\eqno(3.33)
$$
$$
{\partial \tau^{ij}_{AB} \over \partial q^{C}_{k}} +
{\partial \tau^{jk}_{BC} \over \partial q^{A}_{i}} +
{\partial \tau^{ki}_{CA} \over \partial q^{B}_{j}} = 0
{}~~~(i,j,k = 0,1,2).\eqno(3.34)
$$

3B.2 As in Subsection 3A.2 we concentrate first on the functions
$\sigma_{AB}$.
It is not very hard to derive from the closedness conditions the
following consequences:
$$
\sigma_{AB} = \sigma_{BA}\eqno(3.35)
$$
and
$$
{\partial \sigma_{AB} \over \partial q^{C}_{2}} =
{\partial \sigma_{AC} \over \partial q^{B}_{2}} .\eqno(3.36)
$$

3B.3 Let us impose the invariance with respect to the Galilei transformations:
$$
\phi_{R,v,\eta,a} (t,q_{0}) = (t+\eta,Rq_{0}+tv+a).\eqno(3.37)
$$
(We use the same notation as in Subsection 3A.3; in addition we have
$v \in R^{n-1}$).

The lift of these transformation is:
$$
\dot\phi_{R,v,\eta,a} (t,q_{0},q_{1},q_{2},q_{3}) = (t+\eta,Rq_{0}+tv+a,
Rq_{1}+v,Rq_{2},Rq_{3}).\eqno(3.38)
$$

The invariance condition
$$
(\dot\phi_{R,v,\eta,a})^{*} \sigma = \sigma \eqno(3.39)
$$
is equivalent to the fact that the functions
$\sigma_{AB}$,
$\tau^{ij}_{AB}$
and
$\tau_{A}$
do not depend on
$t$,
$q_{0}$
and
$q_{1}$
and are also rotation covariant.

Because
$\sigma_{AB}$
depend only of
$q_{2}$
(see (3.22)) we can use (3.35) and (3.36) to prove that
these function have the generic form:
$$
\sigma_{AB} = \delta_{AB} F(q_{2}^{2}) + 2 q_{2A} q_{2B}
F'(q_{2}^{2}) \eqno(3.40)
$$
for some smooth function
$F$.
The condition of non-degeneracy is
$$
F \not= 0
$$
which will be assumed in the following.

3B.4 Let us define the Lagrangian:
$$
L_{0}(q_{2}) \equiv l(q_{2}^{2})\eqno(3.41)
$$
and take
$l$
such that
$$
F = -2l'\eqno(3.42)
$$
(see (3.40)). Then one can easily prove that:
$$
(\sigma_{L_{0}})_{AB} = \sigma_{AB}.\eqno(3.43)
$$

Moreover
$L_{0}$
verifies a relation of the type (2.49) for every transformation
$
\phi_{R,v,\eta,a}
$
with
$
\Lambda = 0
$.
So, we have:
$$
(\dot\phi_{R,v,\eta,a})^{*} \sigma_{L_{0}} = \sigma_{L_{0}},\eqno(3.44)
$$

The relations above suggest to define an auxilliary
Lagrange-Souriau 2-form:
$$
\sigma' \equiv \sigma - \sigma_{L_{0}}.\eqno(3.45)
$$

Then we have:
$$
(\dot\phi_{R,v,\eta,a})^{*} \sigma' = \sigma'\eqno(3.46)
$$
and
$$
\sigma'_{AB} = 0.
$$

3B.5 We are reduced to a simpler problem, namely the case when:
$$
\sigma_{AB} = 0.\eqno(3.47)
$$

This brings substantial simplifications to the closedness
conditions (3.22)-(3.34). Taking into account the Galilei
invariance (3.46) also we obtain the following solution:
$$
\tau^{00}_{AB} = 0\eqno(3.48)
$$
$$
\tau^{01}_{AB} = - m\delta_{AB}~~~( m \in R)\eqno(3.49)
$$
$$
\tau^{02}_{AB} = \tau^{11}_{AB} =  \kappa\varepsilon_{AB}~~~
( n = 3).\eqno(3.50a)
$$
$$
\tau^{02}_{AB} = \tau^{11}_{AB} = 0~~~( n > 3) \eqno(3.50b)
$$
and
$$
\tau_{A} = - m q_{2A} + \tau^{02}_{AB} q_{3}^{B}.\eqno(3.51)
$$

It remains to note that we have
$
\sigma = \sigma_{L_{1}}
$
where:
$$
L{1}(q_{1},q_{2}) = {1\over 2} m q_{1}^{2} + {1\over 2} \kappa
\varepsilon_{AB} q_{1}^{A} q_{2}^{B}\eqno(3.52)
$$
where the last contibution shows up only for
$n = 3$.

3B.6 Form the last two Subsections it follows that
$
\sigma = \sigma_{L}
$
where:
$$
L = L_{0} + L_{1} = l(q_{2}^{2}) + {1\over 2} m q_{1}^{2} +
{1\over 2} \tau^{02}_{AB} q_{1}^{A} q_{2}^{B}.\eqno(3.53)
$$

3B.7 We turn now to the Poincar\'e invariance; this means that
we keep the invariance with respect to
$
\phi_{R,0,\eta,a}
$
(see (3.37)) but instead of
$
\phi_{1,v,0,0}
$
we consider:
$$
\phi_{\beta,\chi}(t,q_{0}) = (cosh(\chi) t + sinh(\chi)
\beta\cdot q_{0}, q_{0} + \beta [(cosh(\chi) - 1) \beta\cdot
q_{0} + t sinh(\chi)])\eqno(3.54)
$$
for
$
\beta \in S^{2}
$
and
$
\chi \in R
$
and we impose:
$$
(\dot\phi_{\beta,\chi})^{*} \sigma= \sigma.\eqno(3.55)
$$

It is better to consider (3.55) in the infinitesimal form:
$$
L_{X_{\beta}}\sigma = \sigma\eqno(3.56)
$$
where
$
X_{\beta} \in Vect(J^{3}_{1}(S))
$
is the vector field associated to the uni-parametric group action
$
R \ni \chi \mapsto \dot\phi_{\beta,\chi} \in Diff(E)
$
which can be easily computed:
$$
X_{\beta} = \beta\cdot q_{0} {\partial \over \partial t} +
t\beta\cdot {\partial \over \partial q_{0}} + [\beta-(\beta\cdot
q_{1})q_{1}] \cdot {\partial \over \partial q_{1}} -
$$
$$
[(\beta\cdot q_{2})q_{1} + 2(\beta\cdot q_{1}) q_{2}]\cdot
{\partial \over \partial q_{2}} - [3(\beta\cdot q_{1}) q_{2} +
3(\beta\cdot q_{2}) q_{2} + (\beta\cdot q_{3}) q_{1}]\cdot
{\partial \over \partial q_{3}}.\eqno(3.57)
$$

3B.8 We want to obtain from (3.56) another relation on
$
\sigma_{AB}
$
which replaces the independence of
$q_{1}$
from the Galilean case. To this purpose we use an well known relation
$
L_{X} = d i_{X} + i_{X} d
$
and (2.16) to write (3.56) as follows:
$$
d i_{X_{\beta}} \sigma = 0.\eqno(3.58)
$$

{}From this relation we select the coefficient of the differential
form
$
d q^{A}_{3} \wedge \delta q_{0}^{B}
$
and take into account that
$
\beta \in S^{2}
$
is arbitrary; we get:
$$
3\sigma_{AB} q_{1C} + \sigma_{AD} q_{1}^{D} \delta_{BC} +
\sigma_{BD} q_{1}^{D} \delta_{AC} -
{\partial \sigma_{AB} \over \partial q_{1}^{D}}
(\delta^{D}_{C} - q_{1C} q^{D}_{1}) +
{\partial \sigma_{AB} \over \partial q_{2}^{D}}
(q^{D}_{1} q_{2C} + 2q_{2}^{D} q_{1C}) = 0.\eqno(3.59)
$$

Using the rotation covariance and the symmetry property (3.35)
it is easy to see that (3.59) is compatible with the following
structure of
$
\sigma_{AB}
$:
$$
\sigma_{AB} = \delta_{AB} a + q_{1A} q_{1B} b +
q_{2A} q_{2B} c + (q_{1A} q_{2B} + q_{2A} q_{1B}) d.\eqno(3.60)
$$
where
$a,b,c$
and
$d$
are smooth functions of the invariants:
$$
\xi_{1} \equiv q_{1}^{2},~~~\xi_{2} \equiv q_{2}^{2},~~~
\xi_{12} \equiv q_{1}\cdot q_{2}.\eqno(3.61)
$$

We first impose (3.36) and obtain in particular:
$$
c = 2 {\partial a \over \partial \xi_{2}}\eqno(3.62)
$$
and
$$
d = {\partial a \over \partial \xi_{12}}.\eqno(3.63)
$$

Next, we impose (3.59); after some computations we obtain an
equation of the following type:
$$
q_{1C} X_{AB} + q_{2C} Y_{AB} + Z_{A} \delta_{BC} +
Z_{B} \delta_{AC} = 0
$$
which is easily seen to be equvalent to:
$$
X_{AB} = 0\eqno(3.64)
$$
$$
Y_{AB} = 0\eqno(3.65)
$$
and
$$
Z_{A} = 0.\eqno(3.66)
$$

{}From (3.66) we get in particular:
$$
a - (1 - \xi_{1}) b + \xi_{12} d = 0.\eqno(3.67)
$$

Taking into account (3.62) and (3.63) it follows that it is
sufficient to determine the function
$a$.
To ths purpose we consider the coefficient of the tensor
$\delta_{AB}$
in (3.64) and (3.65) and get:
$$
3 a - 2(1 - \xi_{1}) {\partial a \over \partial \xi_{1}}
+ 3 \xi_{12} {\partial a \over \partial \xi_{12}} +
4 \xi_{2} {\partial a \over \partial \xi_{2}} = 0\eqno(3.68)
$$
and
$$
(1 - \xi_{1}) {\partial a \over \partial \xi_{12}} +
2 \xi_{12} {\partial a \over \partial \xi_{2}} = 0.\eqno(3.69)
$$

It is easy to prove that the solution of these equations in the domain
$
\xi \not= 1
$
is of the form
$$
a(\xi_{1},\xi_{2},\xi_{12}) = \vert 1 - \xi_{1}\vert^{-3/2}
F\left({(1 - \xi_{1})\xi_{2} + \xi_{12}^{2} \over (1 - \xi_{1})^{3}}
\right).\eqno(3.70)
$$

If one requires the non-degeneracy one must have
$
F \not\equiv 0
$
which we will suppose in the following. Now the smothness
condition for
$a$
compells us to choose as evolution space on of the following form:
$$
E^{\eta} \equiv \{ (t,q_{0},q_{1},q_{2},q_{3}) \vert sign(1 -
q_{1}^{2}) = \eta \}~~~(\eta = \pm).\eqno(3.71)
$$

{}From (3.62), (3.63) and (3.67) one determines
$b,c$
and
$d$
respectively and this elucidates completely the structure of the
function
$
\sigma_{AB}
$.

3B.9 We use the same trick as in 3B.4. Define the Lagrangian
$L_{0}$
by:
$$
L_{0}(q_{1},q_{2}) \equiv \vert 1 - q_{1}^{2} \vert^{1/2}
l\left({(1 - q^{2}_{1})q^{2}_{2} + (q_{1}\cdot q_{2})^{2} \over
(1 - q^{2}_{1})^{3}} \right)\eqno(3.72)
$$
such that:
$$
F = - 2 l'.\eqno(3.73)
$$

Then it is not very hard to prove that:
$$
(\sigma_{L_{0}})_{AB} = \sigma_{AB}.\eqno(3.74)
$$

Moreover
$L_{0}$
verifies a relation of the type (2.49) with
$
\Lambda = 0
$
for every transformation
of the type
$
\phi_{R,0.\eta,a}
$
and
$
\phi_{\beta,\chi}
$,
so we have:
$$
(\dot\phi_{R,0.\eta,a})^{*} \sigma_{L_{0}} = \sigma_{L_{0}}~~~
(\dot\phi_{\beta,\chi})^{*} \sigma_{L_{0}} = \sigma_{L_{0}}.\eqno(3.75)
$$

If we define the auxiliary Lagrange-Souriau form:
$$
\sigma' \equiv \sigma - \sigma_{L_{0}}\eqno(3.76)
$$
we have:
$$
(\dot\phi_{R,0.\eta,a})^{*} \sigma'= \sigma'~~~
(\dot\phi_{\beta,\chi})^{*} \sigma' = \sigma'.\eqno(3.77)
$$
and
$$
\sigma'_{AB} = 0
$$

3B.10 So again we have succeeded to reduce ourselves to a
simpler problem, namely one with
$$
\sigma_{AB} = 0.\eqno(3.78)
$$

The computations are more easy now and we give only the final result:
$
\sigma = \sigma_{L_{1}}
$
where:
$$
L_{1}(q_{1},q_{2}) = -m \vert 1 - q_{1}^{2} \vert^{1/2} +
{1\over 2} \kappa (q_{0} - t q_{1})\eqno(3.79)
$$
where
$m,\kappa \in R
$
and
$\kappa$
can be non-zero
only for
$n = 2$.

3B.11 Combining the results of the last two Subsection we obtain
the final result: the most general Lagrange-Souriau form for
$r = 3$
with Poincar\'e invariance is
$
\sigma = \sigma_{L}
$
where:
$$
L(q_{1},q_{2}) \equiv \vert 1 - q_{1}^{2} \vert^{1/2}
l\left({(1 - q^{2}_{1})q^{2}_{2} + (q_{1}\cdot q_{2})^{2} \over
(1 - q^{2}_{1})^{3}} \right) + {1\over 2} \kappa (q_{0} - t q_{1}).\eqno(3.80)
$$
\vfil\eject

\pageno=18
\footline{\hss\tenrm\folio\hss}
{\bf 4. The Homogeneous Formalism}

4.1 We will give an alternative formuation of the whole problem
using the so-called homogeneous formalism. This method was used
in [9],[10] for the analysis of the extended objects (with
Poincar\'e and Galilei invariance),

The idea is the following one. Suppose the kinematic manifold
of the system is
$
M:~~(dim(M) = n
$.
Instead of working on
$
J^{r}_{1}(M)
$,
one takes in the general formalism from Section 1
$
S = R \times M
$,
works on
$
J^{r}_{1}(S)
$
and imposes in addition the condition that the reparametrization
invariance is a Noetherian symmetry.

If the local coordinats on
$S$
are
$
(\tau,X^{\mu})~~~(\mu = 1,...,n)
$
then the Lagrange-Souriau 2-form writes:
$$
\sigma = \sigma_{\mu\nu} dX^{\mu}_{r}
\wedge \delta X^{\nu}_{0} +
{1 \over 2} \sum_{i,j=0}^{r-1} \tau_{\mu\nu}^{ij}
\delta X^{\mu}_{i} \wedge \delta X^{\nu}_{j}
+ \tau_{\mu} \delta X^{\mu}_{0} \wedge d\tau\eqno(4.1)
$$
where:
$$
\delta X^{\mu}_{i} \equiv d X^{\mu}_{i} - X^{\mu}_{i+1} d\tau
{}~~~(i = 0,...,r-1).\eqno(4.2)
$$

The action of a reparametrization on
$S$
is constructed as follows: let
$
f \in Diff(R);
$
then we have
$$
\phi_{f}(\tau,X) = (f(\tau),X).\eqno(4.3)
$$

The lift of
$
\phi_{f}
$
to
$
J^{r}_{1}(S)
$
is:
$$
\dot\phi_{f}(\tau,X_{0},...,X_{r}) = (f(\tau),X_{0},g_{1}(\tau,X),...,
g_{r}(\tau,X))\eqno(4.4)
$$
where the functions
$
g_{i}~~~(i = 1,...,r)
$
are recurrsively constructed from:
$$
g_{i+1}^{\mu} = {1 \over f'} {\delta g^{\mu}_{i} \over
\delta \tau}~~~(i = 0,...,r-1).\eqno(4.5)
$$

Here:
$${\delta \over \delta \tau} \equiv {\partial \over \partial \tau}
+ \sum_{i=0}^{r-1} X_{i+1}^{\mu} {\partial \over \partial
X^{\mu}_{i}}\eqno(4.6)
$$
and we have denoted for uniformity:
$$
g_{0}(\tau,X) \equiv X_{0}\eqno(4.7)
$$

Then we postulate reparametrization invariance as follows:
$$
(\dot\phi_{f})^{*} \sigma = \sigma.\eqno(4.8)
$$

4.2 We will illustrate the method for a physically interesting case:
$r = 3$
and
$M$
is the Minkowski space.

First, we list the closedness conditions for
$\sigma$
(see (2.17)-(2.27)):
$$
{\partial \sigma_{\mu\nu} \over \partial X^{\rho}_{3}} = 0.\eqno(4.9)
$$
$$
{\delta\sigma_{\mu\nu} \over \delta \tau} -
{\partial \tau_{\mu} \over \partial X^{\nu}_{3}}
+ \tau^{02}_{\mu\nu} = 0.\eqno(4.10)
$$
$$
\tau^{12}_{\mu\nu} = \sigma_{\nu\mu}.\eqno(4.11)
$$
$$
\tau^{22}_{\mu\nu} = 0.\eqno(4.12)
$$
$$
{\partial \sigma_{\mu\nu} \over \partial X^{\rho}_{0}} -
{\partial \sigma_{\mu\rho} \over \partial X^{\nu}_{0}} +
{\partial \tau^{00}_{\nu\rho} \over \partial X^{\mu}_{3}} = 0.\eqno(4.13)
$$
$$
{\partial \sigma_{\mu\nu} \over \partial X^{\rho}_{1}} +
{\partial \tau^{01}_{\nu\rho} \over \partial X^{\mu}_{3}} = 0 .\eqno(4.14)
$$
$$
{\partial \sigma_{\mu\nu} \over \partial X^{\rho}_{2}} +
{\partial \tau^{02}_{\nu\rho} \over \partial X^{\mu}_{3}} = 0 .\eqno(4.15)
$$
$$
{\partial \tau^{ij}_{\mu\nu} \over \partial X^{\rho}_{3}} = 0~~~
(i,j = 1,2).\eqno(4.16)
$$
$$
{\delta \tau^{00}_{\mu\nu} \over \delta \tau} +
{\partial \tau_{\nu} \over \partial X^{\mu}_{0}} -
{\partial \tau_{\mu} \over \partial X^{\nu}_{0}} = 0.\eqno(4.17)
$$
$$
{\delta \tau^{01}_{\mu\nu} \over \delta \tau} + \tau^{00}_{\mu\nu} -
{\partial \tau_{\mu} \over \partial X^{\nu}_{1}}= 0.\eqno(4.18)
$$
$$
{\delta \tau^{02}_{\mu\nu} \over \delta \tau} + \tau^{01}_{\mu\nu} -
{\partial \tau_{\mu} \over \partial X^{\nu}_{2}}= 0.\eqno(4.19)
$$
$$
{\delta \tau^{ij}_{\mu\nu} \over \delta \tau} +
\tau^{i-1,j}_{\mu\nu} - \tau^{j-1,i}_{\nu\mu} = 0~~~(i,j = 1,2).\eqno(4.20)
$$
$$
{\partial \tau^{ij}_{\mu\nu} \over \partial X^{\rho}_{k}} +
{\partial \tau^{jk}_{\nu\rho} \over \partial X^{\mu}_{i}} +
{\partial \tau^{ki}_{\rho\mu} \over \partial X^{\nu}_{j}} = 0
{}~~~(i,j,k = 0,1,2).\eqno(4.21)
$$

Next, we compute from (4.5) the functions
$g_{i}$
appearing in (4.4); we get:
$$
g_{i}^{\mu} = \sum_{j=1}^{i} g_{i}^{j} X^{\mu}_{j}~~~(i = 1,2,3)\eqno(4.22)
$$
where:
$$
g^{i}_{i} = {1 \over (f')^{i}}~~~(i = 1,2,3)\eqno(4.23)
$$
and
$$
g^{1}_{2} = - {f'' \over  (f')^{3}},~~~
g^{1}_{3} = {3 f'' - f''' f' \over
(f')^{5}},~~~ g^{2}_{3} = - {3 f'' \over (f')^{4}}.\eqno(4.24)
$$

4.3 Finally, we impose Poincar\'e invariance in the target
space. The action of this group on
$S$
is:
$$
\phi_{L,a}(\tau,X_{0}) = (\tau,LX_{0} + a)\eqno(4.25)
$$
with the lift to
$
E \subseteq J^{r}_{1}(S)
$:
$$
\dot\phi_{L,a}(\tau,X_{0},X_{1},X_{2},X_{3}) =
(\tau,LX_{0} + a,LX_{1},
LX_{2},LX_{3}).\eqno(4.26)
$$

The Poincar\'e invariance is formulated as:
$$
(\dot\phi_{L,a})^{*} \sigma = \sigma\eqno(4.27)
$$
and it is easily seen to be equivalent to the independence of the functions
$\sigma_{\mu\nu}$,
$\tau_{\mu\nu}^{ij}$
and
$\tau_{\mu}$
of the variable
$X_{0}$
and also their Lorentz covariance.

4.4 As in Section 3 we concentrate on the analysis of the
functions
$\sigma_{\mu\nu}$.
We have from 4.2 and 4.3 that these functions no not depend on
$X_{3}$
and
$X_{0}$
respectively, so they are functions of
$\tau$,
$X_{1}$
and
$X_{2}$.

We get more conditions from the reparametrization invariance
(4.8). Indeed the coefficient of the differential form
$
d X^{\mu}_{3} \wedge \delta X_{0}^{\nu}
$
in this relation provides:
$$
g^{3}_{3} \sigma_{\mu\nu} \circ \dot\phi_{f} = \sigma_{\mu\nu}
$$
or explicitely (see (4.4) and (4.22)-(4.24)):
$$
{1\over (f')^{3}} \sigma_{\mu\nu}\left(f(\tau),{1\over f'}X_{1},
X_{2}-{f''\over f'}X_{1}\right) = \sigma_{\mu\nu}(\tau,
X_{1},X_{2}).\eqno(4.28)
$$

If we take in (4.28)
$
f(\tau) = \tau + a~~(a \in R)
$
we obtain that
$\sigma_{\mu\nu}$
does not depent on
$\tau$
so it is really a function of
$X_{1}$
and
$X_{2}$.
If we take in (4.28)
$
f(\tau) = \lambda \tau~~~(\lambda \in R_{+})
$
we get an homogeneity property:
$$
\lambda^{-3}\sigma_{\mu\nu}(\lambda^{-1}X_{1},\lambda^{-2}X_{2}) =
\sigma_{\mu\nu}(X_{1},X_{2})~~~(\forall \lambda \in R_{+})\eqno(4.29)
$$
and (4.28) simplifies to
$$
\sigma_{\mu\nu}(X_{1},X_{2}-\alpha X_{1}) = \sigma_{\mu\nu}(X_{1},X_{2})~~~
(\forall \alpha \in R).\eqno(4.30)
$$

To (4.29) and (4.30) one must add relations of the type (3.35)
and (3.36) which are obtained in the same way:
$$
\sigma_{\mu\nu} = \sigma_{\nu\mu}\eqno(4.31)
$$
and
$$
{\partial \sigma_{\mu\nu} \over \partial X^{\rho}_{2}} =
{\partial \sigma_{\mu\rho} \over \partial X^{\nu}_{2}}.\eqno(4.32)
$$

4.5 The analysis of (4.28)-(4.32) proceeds as in section 3.
First, from the Lorentz covariance of
$\sigma_{\mu\nu}$
and (4.30) one has the following generic expression:
$$
\sigma_{\mu\nu} = \delta_{\mu\nu} A + X_{1\mu} X_{1\nu} B +
X_{2\mu} X_{2\nu} C + (X_{1\mu} X_{2\nu} + X_{2\mu} X_{1\nu}) D
\eqno(4.33)
$$
with
$A, B, C$
and
$D$
smooth functions of the invariants:
$$
\zeta_{1} \equiv X_{1}^{2},~~~\zeta_{2} = \equiv X_{2}^{2},~~~
\zeta_{12} \equiv X_{1}\cdot X_{2}.\eqno(4.34)
$$

First one uses (4.32) and obtain:
$$
C = 2 {\partial A \over \partial \zeta_{2}}\eqno(4.35)
$$
$$
D = {\partial A \over \partial \zeta_{12}}\eqno(4.36)
$$
and:
$$
2 {\partial B \over \partial \zeta_{2}} = {\partial^{2} A \over
\partial \zeta_{12}^{2}}.\eqno(4.37)
$$

Next, one uses (4.30) and obtains in particular:
$$
A \circ \phi_{\alpha} = A\eqno(4.38)
$$
$$
B \circ \phi_{\alpha} + \alpha^{2} C \circ \phi_{\alpha}
- 2 \alpha D \circ \phi_{\alpha} = B\eqno(4.39)
$$
where:
$$
\phi_{\alpha}(\zeta_{1},\zeta_{2},\zeta_{12}) \equiv
(\zeta_{1},\zeta_{2}-2\alpha\zeta_{12}+\alpha^{2}\zeta_{1},\zeta_{12}-
\alpha\zeta_{1}).\eqno(4.40)
$$

Finally, from the homogeneity property (4.29) one obtains for
$A$:
$$
\lambda^{-3} A(\lambda^{-2}\zeta_{1},\lambda^{-4}\zeta_{2},
\lambda^{-3}\zeta_{12}) = A(\zeta_{1},\zeta_{2},\zeta_{12})~~~
(\forall \lambda \in R_{+}).\eqno(4.41)
$$

This shows that for
$
\zeta_{1} \not= 1
$,
$A$
must be of the form:
$$
A(\zeta_{1},\zeta_{2},\zeta_{12}) = \vert\zeta_{1}\vert^{-3/2}
A_{1}\left(\vert\zeta_{1}\vert^{-3/2}\zeta_{12},
\vert\zeta_{1}\vert^{-2}\zeta_{2}\right)
$$
for some smooth function
$A_{1}$.
But then (4.38) gives rather easily that in fact we have:
$$
A(\zeta_{1},\zeta_{2},\zeta_{12}) =  \vert\zeta_{1}\vert^{-3/2}
a \left({\zeta_{1}\zeta_{2}-\zeta_{12}^{2} \over \zeta_{1}^{3}}
\right) \eqno(4.42)
$$
for some smooth function
$a$.
The functions
$C$
and
$D$
are obtained from (4.35) and (4.36) respectively:
$$
C(\zeta_{1},\zeta_{2},\zeta_{12}) = 2 \vert\zeta_{1}\vert^{-7/2}
a'\left( {\zeta_{1}\zeta_{2}-\zeta_{12}^{2} \over \zeta_{1}^{3}}
\right)\eqno(4.43)
$$
$$
D(\zeta_{1},\zeta_{2},\zeta_{12}) = -2 \vert\zeta_{1}\vert^{-7/2}
\zeta_{2}\zeta_{1}^{-1}
a'\left( {\zeta_{1}\zeta_{2}-\zeta_{12}^{2} \over \zeta_{1}^{3}}
\right)\eqno(4.44)
$$

It remains to determine
$B$
from (4.37) and (4.39). The result is:
$$
B(\zeta_{1},\zeta_{2},\zeta_{12}) = 2 \vert\zeta_{1}\vert^{-11/2}
\zeta_{12}^{2}
a'\left( {\zeta_{1}\zeta_{2}-\zeta_{12}^{2} \over \zeta_{1}^{3}}
\right) -2\vert\zeta_{1}\vert^{-3/2}
\zeta_{1}^{-1}
a\left( {\zeta_{1}\zeta_{2}-\zeta_{12}^{2} \over \zeta_{1}^{3}}
\right) + b_{0} \vert\zeta_{1}\vert^{-5/2}
\eqno(4.45)
$$
for some
$
b_{0} \in R
$.

We stress once again that (4.42)-(4.45) have been derived only
for
$
\zeta_{1} \not= 1
$.
The condition of non-degeneracy and smoothness compell us to
take as evolution spaces one of the following two possibilities:
$$
E^{\eta} \equiv \{ (\tau,X_{0},X_{1},X_{2},X_{3}) \vert sign(X_{1}^{2})
= \eta \}~~~(\eta = \pm).\eqno(4.46)
$$

4.6 We use now the same trick as in Section 3 namely, we
consider the Lagrangean:
$$
L_{0}(X_{1},X_{2}) \equiv \vert X_{1}^{2}\vert^{1/2}
F\left( {X_{1}^{2}X_{2}^{2}-(X_{1}\cdot X_{2})^{2} \over
(X_{1}^{2})^{3}}\right)\eqno(4.47)
$$
which is well defined on
$E^{\eta}$
and take
$F$
such that:
$$
a = - 2 F'.\eqno(4.48)
$$

Then one can obtain after some computations that:
$$
\sigma_{\mu\nu} - (\sigma_{L_{0}})_{\mu\nu} =
b_{0} \vert \zeta_{1}^{2}\vert^{-5/2} X_{1\mu}X_{1\nu}.\eqno(4.49)
$$

This relation suggests to define as before:
$$
\sigma' \equiv \sigma - \sigma_{L_{0}}.\eqno(4.50)
$$

Then we will have from (4.50):
$$
\sigma'_{\mu\nu} =
b_{0} \vert \zeta_{1}^{2}\vert^{-5/2} X_{1\mu}X_{1\nu}.\eqno(4.51)
$$

It is also not very hard to prove that
$L_{0}$
verifies a relation of the type (2.49) with
$\Lambda = 0$
for every transformation
$\phi_{f}$
and
$\phi_{L,a}$.

So we have:
$$
(\dot\phi_{f})^{*} \sigma_{L_{0}} = \sigma_{L_{0}}\eqno(4.52)
$$
and
$$
(\dot\phi_{L,a})^{*} \sigma_{L_{0}} = \sigma_{L_{0}}\eqno(4.53)
$$

{}From (4.8) and (4.52) it follows that:
$$
(\dot\phi_{f})^{*} \sigma' = \sigma'\eqno(4.54)
$$
and from (4.27) and (4.53) that:
$$
(\dot\phi_{L,a})^{*} \sigma' = \sigma'\eqno(4.55)
$$

4.7 The analysis performed above shows that the starting problem
can be reduced to a simpler one, namely when:
$$
\sigma_{\mu\nu} =
b_{0} \vert \zeta_{1}^{2}\vert^{-5/2} X_{1\mu}X_{1\nu}
$$

But a carefull analysis of the reparametrization invariance
(4.8) shows that in fact we must have
$b_{0} = 0$.
So, in the end in remains to analyse the case:
$$
\sigma_{\mu\nu} = 0.\eqno(4.56)
$$

We outline briefly this analysis. First, one uses the closedness
conditions to prove that
$
\tau_{\mu\nu}^{11} = \tau_{\nu\mu}^{02} = const.
$
Then (4.8) compels us to take:
$$
\tau_{\mu\nu}^{11} = \tau_{\nu\mu}^{02} = 0.\eqno(4.57)
$$

Next, one shows that
$
\tau_{\mu\nu}^{00} = const.
$
Antisymmetry and Lorentz covariance gives:
$$\tau_{\mu\nu}^{00} =  \kappa\varepsilon_{\mu\nu}~~~(n = 2)\eqno(4.58a)
$$
$$
\tau_{\mu\nu}^{00} = 0~~~(n > 2).\eqno(4.58b)
$$

We also obtain from the closedness condition that
$
\tau_{\mu\nu}^{01}
$
depends only on
$X_{1}$,
is symmetric and verifies
$
{\partial  \tau_{\mu\nu}^{01} \over \partial X^{\rho}_{1}} =
\nu \leftrightarrow \rho
$.
Combining with the Lorentz covariance we get:
$$
\tau_{\mu\nu}^{01} = \delta_{\mu\nu} A(X_{1}^{2}) +
2X_{1\mu} X_{1\nu} A'(X_{1}^{2}).
$$
But (4.8) provides a homogeneity property on
$A$
so we end up with
$E = E^{\eta}$
and:
$$
\tau_{\mu\nu}^{01} = \delta_{\mu\nu} m sign(X_{1}^{2})
\vert X_{1}^{2}\vert^{-1/2}  -
X_{1\mu} X_{1\nu} m \vert X_{1}^{2}\vert^{-3/2}
{}~~~( m \in R).\eqno(4.59)
$$.

Finally, Lorentz covariance and the closedness condition determine
$\tau_{\mu}$
to be:
$$
\tau_{\mu} = \tau_{\mu\nu}^{01} X_{2}^{\nu} +
\tau_{\mu\nu}^{00} X_{1}^{\nu}.\eqno(4.60)
$$

It remains to note that
$
\sigma = \sigma_{L_{1}}
$
where:
$$
L_{1}(X_{1}) = - m \vert X_{1}^{2}\vert^{1/2} +
{1\over 2} \tau_{\mu\nu}^{00} X_{0}^{\mu} X_{1}^{\nu}.\eqno(4.61)
$$

So, in the notations of Subsection 4.6,
$
\sigma = \sigma_{L}
$
where the Lagrangian
$L$
has the structure:
$$
L(X_{1},X_{2}) \equiv \vert X_{1}^{2}\vert^{1/2}
F\left( {X_{1}^{2}X_{2}^{2}-(X_{1}\cdot X_{2})^{2} \over
(X_{1}^{2})^{3}}\right) + {1\over 2} \tau_{\mu\nu}^{00}
X_{0}^{\mu} X_{1}^{\nu}\eqno(4.62)
$$
where, of course, the last contribution shows out {\it iff}
$n = 2$.

{\bf Remarks:} 1) The first contribution in (4.62) has been
extensively investigated in [11], where one proves a weaker
result, namely that
$L$
is the generic form of a strictly invariant Lagrangian i.e.
a Lagrangian verifying relations of the type (2.49) with
$\Lambda = 0$
for
$\phi_{f}$
and
$\phi_{L,a}$.
Our analysis proves a much stronger result, namely that every
Lagrangian which is invariant up to a total derivative under
these transformations (i.e. the transformations are Noetherian
symmetries) is equivalent to a Lagrangian of the type (4.62).

2) If we take in (4.62)
$\tau = X_{0}^{0}$
we reobtain (3.80).
\vskip1truecm
{\bf 5.Conclusions}

We have proved that the main results of [1] extends to higher
order Lagrangian theories and we have illustrated the method on
the study of Lagrangian systems with groups of Noetherian
symmetries.

It is plausible that one can extend these results to classical
field theories using ideas from [12],[13].

In this way one would be able to obtain in a systematic way
(using the homogeneous formalism) Lagrangians for extended
objects of the same type as (4.62). Higher-order Lagrangians
for extended objects have been already used in the literature
[14],[15].
\vskip 1truecm

{\bf Acknowledgements} The author wishes to thank
Dr. Olga Krupkova and Dr. D. Krupka for useful discussions.
\vskip1truecm

{\bf References}
\item{1.}
D. R. Grigore, ``Generalized Lagrangian Dynamics and Noetherian
Symmetries'', to appear in Int. Journ. Mod. Phys. A
\item{2.}
J. Klein, ``Espaces Variationels et M\'ecanique'', Ann.
Inst.Fourier 12 (1962) 1-121
\item{3.}
J-M. Souriau, ``Structure des Syst\`emes Dynamiques'', Dunod, Paris, 1970
\item{4.} P. Dedecker, ``Le th\'eoreme de Helmholtz-Cartan pour
une Int\'egrale Simple d'Ordre Superieure'', C. R. Acad. Sci.,
Paris, ser. A 288 (1979) 827-830
\item{5.}
P L. Garcia, J. Munoz, ``On the Geometric Structure of Higher-Order
Variational Calculus'',
Proc. IUTAM-ISIMM Symposium on Modern Developments
in Analytical Mechanics, Torino, 1982, pp. 127-147
\item{6.}
D. Krupka, ``Lepagean Forms in Higher-Order Variational Theory'',
idem, pp. 197-238
\item{7.}
O. Krupkova, ``Lepagean 2-Form in Higher-Order Hamiltonian Mechanics
I. Regularity'', Arch. Math. (Brno) 22 (1986) 97-120
\item{8.}
O. Krupkova, ``Variational Analysis on Fibered Manifolds over
One Dimensional Bases'', Thesis, Opava, 1992
\item{9.}
D. R. Grigore, ``A Derivation of the Nambu-Goto Action from
Invariance Principles'', J. Phys. A 25 (1992) 3797-3811
\item{10.}
D. R. Grigore, ``A Geometric Lagrangian Formalism for Extended
Objects'', to appear in the Proceedings of the
$V^{th}$
Conf. on Diff. Geom., Opava, 1992
\item{11.}
J. Govaerts, ``Relativistic Rigid Particles: Classical Tachions
and Quantum Anomalies'', preprint DTP-92/41
\item{12.}
D. R. Grigore, O. T. Popp, ``On the Lagrange-Souriau Form
in Classical Field Theory'', submitted for publication
\item{13.}
I. M. Anderson, T. Duchamp, ``On the Existence of Global
Variational Principles'', Am. J. Math. 102 (1980) 781-868
\item{14.}
A. M. Polyakov, ``Fine Structure of Strings'', Nucl. Phys. B 268
(1986) 406-412
\item{15.}
H. Kleinert, ``The Membrane Properties of Condensing Strings'',
Phys. Lett. B 174 (1986) 335-338
\bye